\documentclass[twocolumn,aps,pra,showpacs]{revtex4}
\usepackage[latin9]{inputenc}
\usepackage{amsmath}
\usepackage{amssymb}
\usepackage{graphicx}
\usepackage{esint}
\usepackage{amsfonts}
\usepackage{epsfig}

\makeatletter
\@ifundefined{textcolor}{}
{%
 \definecolor{BLACK}{gray}{0}
 \definecolor{WHITE}{gray}{1}
 \definecolor{RED}{rgb}{1,0,0}
 \definecolor{GREEN}{rgb}{0,1,0}
 \definecolor{BLUE}{rgb}{0,0,1}
 \definecolor{CYAN}{cmyk}{1,0,0,0}
 \definecolor{MAGENTA}{cmyk}{0,1,0,0}
 \definecolor{YELLOW}{cmyk}{0,0,1,0}
}

\usepackage{amsfonts}
\usepackage{epsfig}

\makeatother

\begin{document}

\title{Two-axis spin squeezing of two-component BEC via a continuous driving}

\author{Wen Huang, $^{1,2}$ }

\author{Yan-Lei Zhang, $^{1,2}$ }

\author{Chang-Ling Zou, $^{1,2}$}

\email{clzou321@ustc.edu.cn}

\author{Xu-Bo Zou, $^{1,2}$ }

\email{xbz@ustc.edu.cn}

\author{Guang-Can Guo $^{1,2}$ }

\affiliation{$^{1}$ Key Laboratory of Quantum Information, University of Science
and Technology of China, Hefei, 230026, People's Republic of China; }

\affiliation{$^{2}$ Synergetic Innovation Center of Quantum Information \& Quantum
Physics, University of Science and Technology of China, Hefei, Anhui
230026, China}
\begin{abstract}
In two-component BEC, the one-axis twisting Hamiltonian leads to spin
squeezing with the limitation that scales with the number of atoms
as $N^{-\frac{2}{3}}$. We propose a scheme to transform the one-axis
twisting Hamiltonian into a two-axis twisting Hamiltonian, resulting
in enhanced spin squeezing $\propto N^{-1}$ approaching the Heisenberg
limit. Instead of pulse sequences, only one continuous driving field
is required to realizing such transforming, thus the scheme is promising
for experiment realizations, to an one-axis twisting Hamiltonian.
Quantum information processing and quantum metrology may benefit from
this method in the future.
\end{abstract}


\maketitle
{\em Introduction.} Squeezed spin states (SSSs)~\cite{Kitagawa93,Wineland94,Ma11},
whose concept was firstly established by Kitagawa and Ueda~\cite{Kitagawa93},
are entangled quantum states of an ensemble of spin systems. The SSS
attracted considerable attention due to their significant roles in
studying many-particle entanglement~\cite{Sorensen01,Bigelow01,Sorensen01L,Amico08,Horodecki09,Guhne09}
and applications for high-precision measurements~\cite{Wineland94,Wineland92,Bollinger96,Polzik08,Cronin09,Agarwal96,Meiser08,Andre04}.
In the original proposal ~\cite{Kitagawa93}, there are two distinguished
ways to produce SSS, one interaction in the form as $\chi J_{x}^{2}$ is known as one-axis twisting (OAT), the other one in the form
as $\chi(J_{+}^{2}-J_{-}^{2})$ is known as two-axis twisting
(TAT). The OAT scheme just can reduce the noise limit to the scale
as $N^{-\frac{2}{3}}$ where $N$ is atom number, while the TAT can
produce the SSS with the squeezing parameter scaling with $N^{-1}$ \cite{Kitagawa93}.
Both in theory and experiment, most schemes can only produce effective OAT-type spin-spin
interactions, such as direct atom collisions in Bose-Einstein condensates (BEC)~\cite{Gross10,Riedel10,Julia12},
indirect spin-spin interaction by quantum nondemolition measurement~\cite{Chaudhury07,Takano09,Inoue13,Leroux12,Julsgaard01,Louchet10}
and cavity feedback~\cite{Schleier10,Leroux10}.

The two-component BEC is a very promising system for OAT SSSs \cite{Poulsen01,Raghavan01,Jenkins02},
and it has been demonstrated in experiments recently \cite{Orzel01,Esteve08,Gross10,Riedel10}.
It holds two main advantages, including the considerable long coherence
time and the strong atom-atom interaction, is very potential for future
applications. Therefore, various efforts are dedicated to realizing the
TAT type Hamiltonian to enhance the squeezing in such system \cite{Helmerson01,Liu11,Duan13,Zhang14}.
One of the proposals~\cite{Liu11} transforms an OAT Hamiltonian
into an effective TAT Hamiltonian by applying a large number of repeated
Rabi pulses, which would be sensitive to the accumulation of control
errors. In another scheme~\cite{Duan13}, one or two global rotation
pulses are applied at an appropriate evolution time and with optimized
rotation angles, which reduces the number of pulses greatly, but requires
a long evolution time to achieve the optimal squeezing and the control
pulse is spin number dependent.

In this paper, we propose a scheme to transform the OAT into the
effective TAT spin squeezing in BEC by continuous coherent driving.
Under the driving, the spin state is rotating along the direction
perpendicular to the twisting axis, then generate the effect Hamiltonian
as mixed OAT and TAT. By carefully choosing and tuning the amplitude
and and frequency of the driving field, pure TAT can be realized and
a Heisenberg limited noise reduction $\propto N^{-1}$ is obtained.
Compared with the previous scheme~\cite{Liu11}, our proposal uses
a continuous field instead of pulse sequences, which is more friendly
for experiments. What's more, our scheme is spin number independent
and needs a shorter evolution time compared with~\cite{Duan13}.
The principle of continuous driving transformed the OAT to the TAT
can also be applied to other systems, such as the cavity feedback
~\cite{Schleier10,Leroux10} and spin state dependent geometry phase~\cite{Yan14} induced OAT.

{\em Theoretical Model.} According to Refs.~\cite{Gross10,Riedel10},
the two-component BEC with a coherent driving can be described by
the following Hamiltonian
\begin{equation}
H=\chi J_{x}^{2}+\Omega(t)J_{z}.\label{Hamiltonian}
\end{equation}
Here $J_{\mu}=\sum_{k=1}^{N}\sigma_{\mu}^{k}/2$ in terms of the Pauli
matrices $\sigma_{\mu}^{k}$ $(\mu=x,y,z)$ is the collective angular
momentum operator for the spin ensemble consisting of $N$ atoms. The
first term of the Hamiltonian is the OAT induced by atom-atom collisions,
with $\chi$ the nonlinear interaction strength. The second term is
the external classical laser driving with magnetic field along the
$z$-axis. For the continuous driving, we assume $\Omega(t)=g\cos(\omega t)$,
where $g$ and $\omega$ are the strength and frequency of the driving
field, respectively.

Transform the Hamiltonian~(\ref{Hamiltonian}) into the interaction
representation, we get
\begin{eqnarray}
H_{I} & = & e^{i\int_{0}^{t}\Omega(\tau)J_{z}d\tau}\chi J_{x}^{2}e^{-i\int_{0}^{t}\Omega(\tau)J_{z}d\tau}\nonumber \\
 & = & \frac{\chi}{4}(e^{2igC}J_{+}^{2}+e^{-2igC}J_{-}^{2}+J_{+}J_{-}+J_{-}J_{+}),\label{H_I}
\end{eqnarray}
where $C=\int_{0}^{t}\cos(\omega t)dt=\frac{\sin(\omega t)}{\omega}$
and $J_{\pm}=J_{x}\pm iJ_{y}$. According to the Jacobi-Anger expansion
$e^{iz\sin\theta}=\sum_{n=-\infty}^{\infty}{\cal J}_{n}(z)e^{in\theta}$
where ${\cal J}_{n}(z)$ is the $n$-th Bessel function of the first
kind, the terms in Eq. (\ref{H_I}) can be expanded as
\begin{equation}
e^{\pm2igC}=e^{\pm i\frac{2g}{\omega}\sin(\omega t)}=\sum_{n=-\infty}^{\infty}{\cal J}_{n}(\pm\frac{2g}{\omega})e^{in\omega t}.\label{eq:Jacobi-Anger}
\end{equation}
When $\omega$ is quite large ($\omega\gg N\chi$), the high-order
terms with $n\neq0$ are neglected due to the rotating wave approximation.
Then, the Hamiltonian becomes
\begin{equation}
H_{I}'\simeq\frac{\chi}{2}[(A+1)J_{x}^{2}-(A-1)J_{y}^{2}],\label{(H_I')}
\end{equation}
where the constant $A={\cal J}_{0}(\frac{2g}{\omega})$. Therefore,
the external driving field leads to the twisting effect along both
$x$ and $y$ directions. This can be interpreted intuitively as the
rotation of spins perpendicular to the axis of OAT ($x$-axis) diverted
the twisting axis.
\begin{figure}
\includegraphics[width=0.95\columnwidth,height=2.8in]{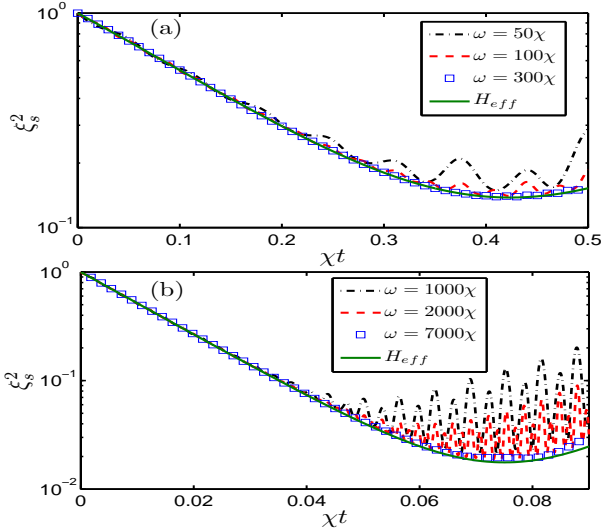}
\protect\protect\caption{(Color online) (a) Spin squeezing parameter as a function of the evolution
time for different frequencies of the driving field with the number
of atoms $N=10$. The frequencies are $\omega=50\chi$ (dot-dashed
dark line), $\omega=100\chi$ (dashed red line), $\omega=300\chi$
(blue squares) and $H_{eff}$ (solid blue line). (b) Same as (a) except
for $N=100$ with the frequencies corresponding to $\omega=1000\chi$
(dot-dashed dark line), $\omega=2000\chi$ (dashed red line), $\omega=7000\chi$
(blue squares) and $H_{eff}$ (solid blue line). $\frac{g}{\omega}=0.906$
in both (a) and (b).}

\label{Fig1}
\end{figure}

Rewriting the Hamiltonian by adding a constant $\frac{\chi}{2}(A-1)J^{2}$
(which is conserved during the dynamics), we obtain a mixture of an
OAT Hamiltonian and a TAT Hamiltonian as
\begin{equation}
H_{I}''=\frac{\chi}{2}(3A-1)J_{x}^{2}+\frac{\chi}{2}(1-A)(J_{x}^{2}-J_{z}^{2}).\label{(H_I'')}
\end{equation}
Tune the values of $g$ and $\omega$ to be $\frac{g}{\omega}=0.906$,
then $A={\cal J}_{0}(\frac{2g}{\omega})=\frac{1}{3}$ and the effective
Hamiltonian of the system becomes
\begin{equation}
H_{eff}=\frac{\chi}{3}(J_{x}^{2}-J_{z}^{2}).\label{(Heff)}
\end{equation}
Obviously, $H_{eff}$ exhibits the well-known TAT Hamiltonian. Alternatively,
we can also adjust the parameters to satisfy ${\cal J}_{0}(\frac{2g}{\omega})=-\frac{1}{3}$,
then we obtain another TAT Hamiltonian
\begin{equation}
H_{eff}'=\frac{\chi}{3}(J_{y}^{2}-J_{z}^{2}).\label{(Heff')}
\end{equation}

Therefore, the OAT Hamiltonian can be transformed into the TAT Hamiltonian
by tuning the amplitude and frequency of the driving field. Similar
ideas have been studied by Law et al. \cite{Law}, where the underlying
physics is the same with the continuous driving method studied
here. In that work, a steady field are applied for coherent controlling
of the SSS, which is consist with our model with $\omega=0$, effectively
generate a mixture of OAT and TAT. It's worth noting that the effective
nonlinear interaction strength reduces to $1/3$, which is due to
the cancellation of part of spin squeezing when rotating of the squeezing
direction.

{\em Numerical results.} To verify our idea above, we study
the spin squeezing numerically by solving the evolution of spin state.
The initial state is chosen to be a coherent spin state (CSS)~\cite{Kitagawa93}
along the $y$ axis, which is $|\varphi(0)\rangle=2^{-J}\sum_{k=0}^{2J}i^{k}\sqrt{(2J)!/(k)!(2J-k)!}|J,J-k\rangle$
satisfying $J_{y}|\varphi(0)\rangle=J|\varphi(0)\rangle$ with $J=N/2$,
where $|J,k\rangle$ are the eigenstates of $J_{z}$. We choose squeezing
parameter $\xi_{s}^{2}\equiv4min(\triangle J_{\overrightarrow{n}_{\perp}})^{2}/N$~\cite{Kitagawa93}
to quantify the squeezing, where $\overrightarrow{n}_{\bot}$ refers
to the direction perpendicular to the mean spin direction and the
minimization is taken over all such directions.
\begin{figure}
\includegraphics[width=0.9\columnwidth,height=2.6in]{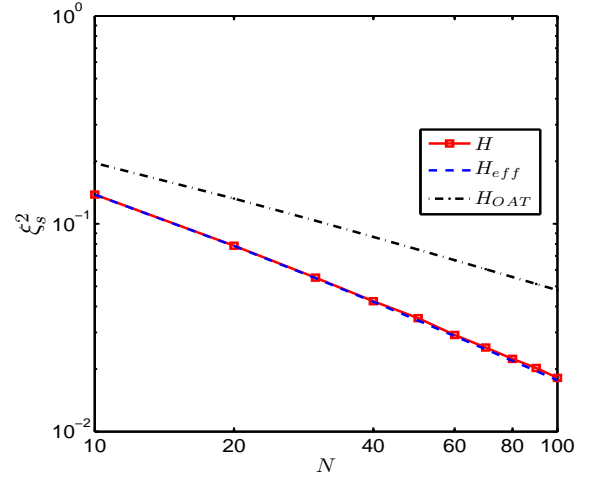}
\protect\protect\caption{(Color online) The optimal spin squeezing plotted against the number
of atoms $N$ for dynamics generated by $H$ (square red line), $H_{eff}$
(dashed blue line) and $H_{OAT}$ (dotted-dashed line).}

\label{Fig2}
\end{figure}

In Fig. 1, we plot the spin squeezing parameter as a function of the
evolution time for different driving frequency $\omega$ but fixed
the ratio that $\frac{g}{\omega}=0.906$ to obtain optimized TAT.
The results for both the $N=10$ (a) and $N=100$ (b) are agrees well
with the effective TAT Hamiltonian {[}Eq. \ref{(Heff)}{]}. There
are fast oscillations of the $\xi_{s}^{2}$ for small $\omega$, which
is attributed to the high-order terms in the Jacobi-Anger expansion
{[}Eq. \ref{eq:Jacobi-Anger}{]} when $\omega/\chi\gg N$ is not satisfied.
For example, the oscillation period for $\omega=50\chi$ is about
$T=0.06/\chi$, corresponding to $\omega T\approx\pi$ which consist
with the period of high order terms. Therefore, higher frequency $\omega$
is favorable for larger number of atoms. In addition, we find that
when the number of atoms $N$ increases, it needs a shorter time to
reach the optimal squeezing, and the time is already much shorter
than the scheme~\cite{Duan13}.
\begin{figure}
\includegraphics[width=0.9\columnwidth,height=3.6in]{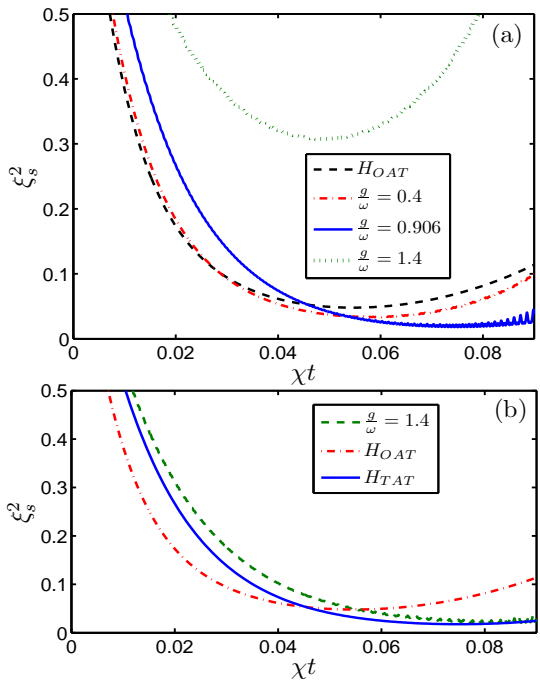}
\protect\protect\caption{(Color online) (a) Spin squeezing parameter as a function of the evolution
time for different $\frac{g}{\omega}$ with $N=100$. The curves are
$\frac{g}{\omega}=0.4$ (dotted-dashed red line), $\frac{g}{\omega}=0.906$
(solid blue line), $\frac{g}{\omega}=1.4$ (dotted green line) and
$H_{OAT}$ (dashed dark line). The initial state is along the $y$
axis. (b) Spin squeezing parameter as a function of the evolution
time with $\frac{g}{\omega}=1.4$ (dashed green line), $H_{OAT}$
(dot-dashed red line) and $H_{TAT}$ (solid blue line) with $N=100$.
The initial state is along the $x$ axis.}

\label{Fig3}
\end{figure}

Next, we investigate how the optimal squeezing of $H$ without approximation
scales with $N$. We plot the optimal spin squeezing (minimum value
of $\xi_{s}^{2}$) as a function of the number of atoms in Fig. 2.
The red solid line corresponding to $H$ shows the optimal spin squeezing
parameter $\xi_{s}^{2}\propto N^{-1}$ which is the well-known Heisenberg
limited noise reduction, and it agrees well with $H_{eff}$ (the blue
dashed line). For comparison, we also present the $N^{-\frac{2}{3}}$
scaling of the OAT Hamiltonian $H_{OAT}=\chi J_{x}^{2}$.
\begin{figure}
\includegraphics[width=0.9\columnwidth,height=3.6in]{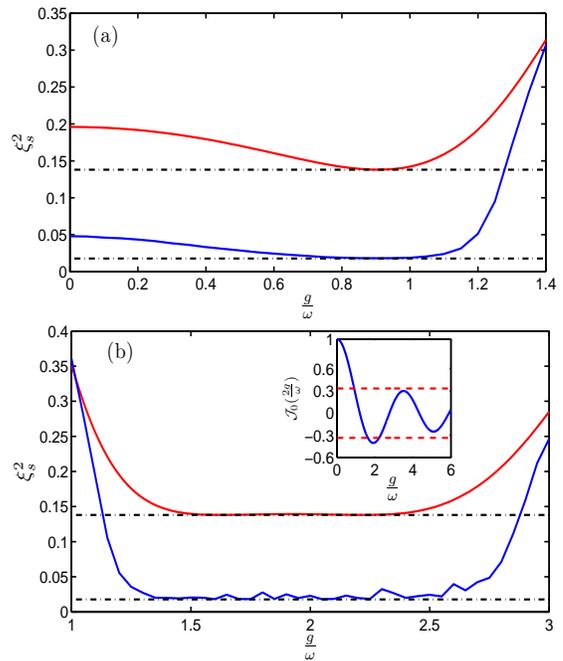}
\protect\protect\caption{(Color online) (a) The optimal spin squeezing as a function of $\frac{g}{\omega}$
with $N=10$ (solid red line), $N=100$ (solid blue line). The two
horizontal lines correspond to the optimal squeezing of the TAT Hamiltonian
$\xi_{s}^{2}=0.1381$ ($N=10$) and $\xi_{s}^{2}=0.0177$ ($N=100$).
The initial state is along the $y$ axis. (b) The same as (a) except
for the initial state along the $x$ axis. The inset shows the Bessel function
${\cal J}_{0}(2g/\omega)$ varying as $g/\omega$, and the horizontal lines indicate ${\cal J}_{0}(2g/\omega)= 1/3, -1/3$ respectively.}

\label{Fig4}
\end{figure}

Although the optimal TAT should satisfy $\frac{g}{\omega}=0.906$,
we could expect the enhanced spin squeezing by continuous driving
field is robust against imperfection parameters, since the external
driving field could lead to mixture of OAT and TAT effectively {[}Eq.~\ref{(H_I'')}{]}.
In Fig. 3(a), we show the squeezing parameter as a function of the
evolution time for different $\frac{g}{\omega}$ with $N=100$, the
dynamics under $H_{OAT}$ is also presented for comparison. We can
find that at $\frac{g}{\omega}=0.4$ the optimal squeezing generated
by $H$ is 0.02805. It is better than that generated by $H_{OAT}$
(0.0479) while worse than that generated by $H_{TAT}$ (0.0177). But
at $\frac{g}{\omega}=1.4$, the squeezing is even worse than $H_{OAT}$,
which is owing to ${\cal J}_{0}(2.8)$ approaching $-\frac{1}{3}$,
then the Hamiltonian $H$ is close to $H_{eff}'=\frac{\chi}{3}(J_{y}^{2}-J_{z}^{2})$.
If we change the initial state correspondingly along $x$-axis, which
is $|\varphi(0)\rangle=2^{-J}\sum_{k=0}^{2J}\sqrt{\frac{(2J)!}{(k)!(2J-k)!}}|J,J-k\rangle$
satisfying $J_{x}|\varphi(0)\rangle=J|\varphi(0)\rangle$, the effect
of this Hamiltonian approaches the idea TAT, as shown in Fig. 3(b).
Therefore, our scheme can always enhance the OAT Hamiltonian to achieve
better SSS even though the achievable $\frac{g}{\omega}$ is deviated
from optimal value.

Finally, we plot the optimal spin squeezing parameter of the Hamiltonian
$H$ as a function of $\frac{g}{\omega}$ in Fig. 4(a) with the initial
state being a CSS along the $y$ axis and in Fig. 4(b) with the initial
state being a CSS along the $x$ axis. In Fig. 4(a), the minimum value
equals to the optimal squeezing of the TAT Hamiltonian appears at
$\frac{g}{\omega}\simeq0.906$, which agrees with the optimal condition.
The rapid growth of $\xi_{s}^{2}$ for $\frac{g}{\omega}>1.2$ is
due to the Hamiltonian changing to $H_{eff}'$. In Fig. 4(b), it shows
a section of gentle variance which almost equals to the optimal squeezing
of the TAT Hamiltonian. There are two points $\frac{g}{\omega}\simeq1.626,\ 2.221$
which make ${\cal J}_{0}(\frac{2g}{\omega})=-\frac{1}{3}$, and the
minimum value of the Bessel function ${\cal J}_{0}(z)$ between the
two points is about $-0.4027$ [inset of Fig. 4(b)] which has no large variance comparing
with $-1/3$. Thus, there is a quite large range approaching the TAT
squeezing, which is favorable for experiments.

{\em Conclusions.} We have proposed a scheme to transform an OAT
Hamiltonian into a TAT type by applying a continuous driving field.
We find that a TAT Hamiltonian can be obtained by tuning the ratio
of the driving field amplitude to the frequency, and even though at
other more achievable values of $\frac{g}{\omega}$, the squeezing
performance of our scheme is more better than the OAT scheme. Compared
with the previous proposals~\cite{Liu11}\cite{Duan13}, our scheme
is more friendly for experiments and faster. Since the continuous
driving field can be manipulated relatively easily, we believe it
is realizable with current techniques as reported in Ref.~\cite{Gross10,Riedel10}.

{\em Acknowledgments.} This work was supported by National Fundamental
Research Program, National Natural Science Foundation of China (No.
11274295, 2011cba00200) and Doctor Foundation of Education Ministry
of China (No. 20113402110059).

\end{document}